\documentclass{svjour2}
\usepackage{graphicx}

\def\beq{\begin{equation}}
\def\eeq{\end{equation}}
\def\rmd{{\rm d}}

\def\fl{}

\journalname{General Relativity and Gravitation}

\begin{document}

\title{Spin-geodesic deviations in the Schwarzschild spacetime}

\author{Donato Bini \and
        Andrea Geralico \and
        Robert T. Jantzen 
}

\institute{
              Donato Bini 
              \at
              Istituto per le Applicazioni del Calcolo ``M. Picone,'' CNR, I-00185 Rome, Italy\\
              ICRA, University of Rome ``La Sapienza,'' I--00185 Rome, Italy\\
              INFN - Sezione di Firenze, Polo Scientifico, Via Sansone 1, I--50019, Sesto Fiorentino (FI), Italy\\
              \email{binid@icra.it} 
\and
              Andrea Geralico 
              \at
              Physics Department and ICRA, University of Rome ``La Sapienza,'' I--00185 Rome, Italy\\
              \email{geralico@icra.it}   
\and
              Robert T. Jantzen
              \at
              Department of Mathematical Sciences, Villanova University, Villanova, PA 19085, USA\\
               \email{robert.jantzen@villanova.edu}   
}

\date{Received: date / Accepted: date / Version: date}

\maketitle

\begin{abstract}
The deviation of the path of a spinning particle from a circular geodesic in the Schwarzschild spacetime
is studied by an extension of the idea of geodesic deviation. Within the Mathisson-Papapetrou-Dixon model and assuming the spin parameter to be sufficiently small so that it makes sense to linearize the equations of motion in the spin variables as well as in the geodesic deviation, the spin-curvature force adds an additional driving term to the second order system of linear ordinary differential equations satisfied by nearby geodesics. 
Choosing initial conditions for geodesic motion leads to solutions for which the deviations are entirely due to the spin-curvature force, and one finds that the spinning particle position for a given fixed total spin oscillates roughly within an ellipse in the plane perpendicular to the motion, while the azimuthal motion undergoes similar oscillations plus an additional secular drift which varies with spin orientation.

\end{abstract}

\section{Introduction}

The  equations of motion  for a spinning test particle in a given gravitational background resulting in the so called
Mathisson-Papapetrou model \cite{math37,papa51} are rather difficult to solve even for highly symmetric  spacetimes.
Trying to obtain solutions which describe particle motion along Killing trajectories (like circular orbits, for example) is in general too restrictive to yield nontrivial solutions since the forces associated with the nongeodesic motion complicate matters \cite{bdfg2004,bdfgj2005,bgj2006}. 

However,  implicit in the model is the requirement that the spin structure of the particle should produce very small deviations from geodesic motion in the sense that the length scale naturally associated with the spin should be very small compared to the one associated with the curvature tensor of the spacetime itself. If this were not the case, one would have to take into account the particle back reaction on the spacetime metric, i.e., approach the problem from a completely different point of view.
It therefore seems reasonable to make some progress towards understanding the consequences of the model by introducing  the condition of ``small spin" from the very beginning, resulting in a simplified set of linearized differential equations which can be easily integrated as shown below.
This approach allows an analytic discussion of the problem in complete generality in this limit, which can be compared with
the previous numerical studies of the full nonlinear equations \cite{maeda,hartl1,hartl2}.
   
One finds that the general solution of the linearized equations of motion corresponds to an adjustment of constants of the motion for the stable oscillations of the nearby geodesics about a new equilibrium circular orbit due to the additional radial component of the spin force arising from the constant spin component out of the equatorial plane. This new equilibrium orbit has a slightly adjusted azimuthal velocity to balance this radial spin force and a slightly adjusted radius. 
The spin component in the equatorial plane (boosted to the local rest space of the world line) undergoes a simple  rotation corresponding to parallel transport along the original circular geodesic and induces an oscillation of the particle path in and out of the equatorial plane adding to the geodesic oscillations already present.

Let us briefly recall the Mathisson and Papapetrou model, which uses a ``center of mass line'' for the spinning particle to perform a multipole reduction. Let $U^\alpha=\rmd x^\alpha /\rmd\tau$ be the timelike unit tangent vector to this world line, parametrized by the proper time $\tau$.
The equations of motion are
\begin{eqnarray}
\label{papcoreqs1}
\frac{DP^{\mu}}{\rmd \tau}&=&-\frac12R^{\mu}{}_{\nu\alpha\beta}U^{\nu}S^{\alpha\beta}\equiv F^{\rm (spin)}{}^{\mu}\,, \\
\label{papcoreqs2}
\frac{DS^{\mu\nu}}{\rmd \tau}&=&P^{\mu}U^{\nu}-P^{\nu}U^{\mu}\,,
\end{eqnarray}
where $P^{\mu}$ is the total 4-momentum of the particle and  $S^{\mu\nu}$ is its antisymmetric (intrinsic angular momentum) spin tensor, both fields defined only along this center of mass world line. 
This system of 10 equations evolve $P$ and $S$ along this world line but contain 13 unknown quantities: $U$ (3), $P$ (4), $S$ (6).
In order for the model to be consistent, the above equations of motion must be completed 
with the Dixon-Tulczyjew \cite{tulc59,dixon64,dixon69,dixon70,dixon73,dixon74} supplementary conditions  
\beq
\label{Tconds}
S^{\mu\nu}P_\nu=0\,,
\eeq
under which the trajectory of the extended body is determined by the position of the center of mass of the body itself.
Other possible choices of supplementary conditions will not be considered here. The present model with these supplementary conditions will be referred to as the Mathisson-Papapetrou-Dixon model.

It is worth mentioning that there exists a wide literature in which the Lagrangian formulation is used to study spinning particle motion, including the interaction with external fields as well (see e.g. \cite{grassberger,cognola}).
Furthermore, it is possible to obtain the Mathisson-Papapetrou-Dixon equations (\ref{papcoreqs1})--(\ref{Tconds}) directly from the Dirac equation by taking a proper limit through the WKB approximation, as shown e.g. in \cite{rudiger}. 

Contracting both sides of Eq.~(\ref{papcoreqs2}) with $U_\nu$, one obtains the following expression linking the total 4-momentum to the 4-velocity and spin tensor derivative
\begin{equation}
\label{SchwPs}
P^{\mu}=mU^\mu -U_\nu \frac{DS^{\mu\nu}}{\rmd \tau}\,,
\end{equation}
where the quantity $m=-U_\mu P^\mu$ is a convenient definition which reduces to the ordinary (constant) particle mass when the spin vanishes. Contracting this momentum relation instead with $P_\mu$ yields
\begin{equation}
\label{Pnorm}
 P_\mu P^\mu = -m^2 + \frac{DS^{\mu\nu}}{\rmd \tau} U_\nu  \frac{DS_\mu{}^{\beta}}{\rmd \tau} U_\beta\,.
\end{equation}

By contracting both sides of Eq.~(\ref{papcoreqs2}) with $-P_\nu$ and using the conditions (\ref{Tconds}), re-expressing the left hand side using the product rule leads to
\begin{equation}
\label{keyapprox}
S^{\mu\nu}\frac{DP_{\nu}}{\rmd \tau}  
= m P^\mu -P_\nu P^\nu U^\mu\,.
\end{equation}
The left hand side is at least of second order in the spin due to Eq.~(\ref{papcoreqs1}), while the second term on the right hand side has a factor containing a second order term in the spin as from Eq.~(\ref{Pnorm}). 
Neglecting these second order terms leaves the relation $0 \approx m(P^\mu - m U^\mu)$ valid to first order in the spin,
which implies that $P$ and $U$ are parallel in this limit: $P^{\mu}\approx mU^\mu$.

The projection of the spin tensor into the local rest space of $U$ (i.e., the subspace of the tangent space orthogonal to $U$) defines the spin vector by spatial duality
\beq
\label{spinvec}
S^\beta={\textstyle\frac12} \eta_\alpha{}^{\beta\gamma\delta}U^\alpha S_{\gamma\delta}=U^\alpha[{}^*S]_\alpha{}^\beta\,,\qquad
S_{\mu\nu}=\eta_{\alpha\beta\mu\nu}U^\alpha S^\beta\,,
\eeq
where $\eta_{\alpha\beta\gamma\delta}=\sqrt{-g} \epsilon_{\alpha\beta\gamma\delta}$ is the unit volume 4-form and $\epsilon_{\alpha\beta\gamma\delta}$ ($\epsilon_{0123}=1$) is the Levi-Civita alternating symbol. 
It is useful to introduce the magnitude $s\ge0$ of the spin vector
\beq
\label{sinv}
s^2=S^\beta S_\beta = \frac12 S_{\mu\nu}S^{\mu\nu}\,, 
\eeq
which is constant along the trajectory of a spinning particle due to Eq.~(\ref{papcoreqs2}). Then $S = s N$, where $N$ is the unit direction of the spin vector.
 
The requirement which is essential to the validity of the Mathisson-Papapetrou-Dixon model and the test particle approach is that the characteristic length scale $|s|/m$ associated with the particle's internal structure be small compared to the natural length scale $M$ (say) associated with the background field \cite{mol}. Hence the following condition must be assumed: $|s|/(mM)\ll 1$. This leads us to consider only the  terms of first order in the spin in Eqs.~(\ref{papcoreqs1}) and (\ref{papcoreqs2}) and to neglect the higher order terms. 
In practice this approximation is usually adopted if not before then after solving the equations of motion in order that the solution make physical sense, describing a test particle with a spin which is in fact small compared with the background geometry.

The set of equations (\ref{papcoreqs1}) and (\ref{papcoreqs2}) then reduces to
\begin{eqnarray}
\label{pap_small_spin}
\frac{D(mU^{\mu})}{\rmd \tau} &=& -\frac12R^{\mu}{}_{\nu\alpha\beta}U^{\nu}S^{\alpha\beta}+O(2)\,, \\
\frac{DS^{\mu\nu}}{\rmd \tau} &=& O(2)\,,
\end{eqnarray}
implying that $\rmd m/\rmd\tau=0$ (since the right hand side of Eq.~(\ref{pap_small_spin})$_1$ is orthogonal to $U$), i.e., the mass of the particle remains constant along the path, and the spin tensor is parallel transported along the path.
Expressing the spin tensor in terms of the spin vector
and introducing the (right) dual $R^*$ of the Riemann tensor, the previous equations can be rewritten in the more convenient form
\begin{eqnarray}
\label{papcoreqs1Iord}
m\frac{DU^{\mu}}{\rmd \tau}&=&-s{\mathcal H}(U)^{\mu\rho} N_{\rho} = F^{\rm (spin)}{}^\mu\,,\\
\label{papcoreqs2Iord}
s\frac{DN^{\mu}}{\rmd \tau}&=&0\,,
\end{eqnarray}
to first order in spin (dropping the $O(2)$ notation from now on), where 
${\mathcal H}(U)^{\mu\rho}= -[R^*]^{\mu\nu\rho\sigma}U_{\nu}U_{\sigma}$
is a trace-free spatial tensor orthogonal to $U$ referred to as the magnetic part of the Riemann tensor with respect to $U$, which in vacuum is also a symmetric tensor.
At this point the total spin $s$ appears in the equations only through the specific spin angular momentum ratio $s/m$. 

Consider a pair of world lines with approximately the same initial data,
one a geodesic with 4-velocity $U_{(\rm geo)}$, the other a world line of a spinning particle which deviates from the geodesic because of the combined effects of geodesic deviation and the spin-curvature coupling, with 4-velocity $U(\tau)$. Solutions of the equation of motion to first order in the spin can then be found in the general form
\beq
\label{Uspindef}
U=U_{\rm (geo)} +\left(\frac{s}{m}\right)Y
\,.
\eeq
Moreover, the normalization condition $U\cdot U=-1$ (to first order in $s$) constrains $Y$ so that
\beq
Y\cdot U_{\rm (geo)}=0\,,
\eeq
synchronizing the proper times to first order so that $\tau$ can be used unambiguously for that single proper time parametrization of both world lines.
Furthermore, this implies that to first order in the spin, the spin orientation vector $N$ is parallely propagated along $U_{\rm (geo)}$.
When the background spacetime admits a nontrivial Killing-Yano tensor $f$,
such a vector $N$ can be directly obtained from $f$, as shown by Marck \cite{Marck1,Marck2}. 

Finally when the background spacetime has Killing vectors, there are conserved quantities associated with particle motion. For example in the case of stationary axisymmetric spacetimes with coordinates adapted to the spacetime symmetries,  $\xi=\partial_t$ is the timelike Killing vector and $\eta=\partial_\phi$ is the azimuthal Killing vector and the corresponding conserved quantities are the total energy $E$ and the angular momentum $J$, namely
\begin{eqnarray}
\label{totalenergy}
\fl\qquad
E&=&-\xi_\alpha P^\alpha +\frac12 S^{\alpha\beta}\nabla_{\beta}\xi_\alpha\,,\qquad
J=\eta_\alpha P^\alpha -\frac12 S^{\alpha\beta}\nabla_{\beta}\eta_\alpha\,.
\end{eqnarray} 

\section{Spinning particles in arbitrary motion in the Schwarzschild spacetime}

Consider the case of the Schwarzschild spacetime with metric 
\beq\fl\quad 
\label{metric}
\rmd  s^2 = -\left(1-\frac{2M}r\right)\rmd t^2 + \left(1-\frac{2M}r\right)^{-1} \rmd r^2 
+ r^2 (\rmd \theta^2 +\sin^2 \theta \rmd \phi^2)\,,
\eeq
and introduce the usual orthonormal frame adapted to the static observers following the time lines
\beq\fl\quad 
\label{frame}
e_{\hat t}=(1-2M/r)^{-1/2}\partial_t, \,
e_{\hat r}=(1-2M/r)^{1/2}\partial_r, \,
e_{\hat \theta}=\frac{1}{r}\partial_\theta, \,
e_{\hat \phi}=\frac{1}{r\sin \theta}\partial_\phi ,
\eeq
where $\{\partial_t, \partial_r, \partial_\theta, \partial_\phi\}$ is the coordinate frame.

We need first to choose a geodesic reference world line, say with 4-velocity
\beq
U_{\rm (geo)}=\gamma_{\rm (geo)} (e_{\hat t} + \nu_{\rm (geo)}^{\hat a} e_{\hat a})\,. 
\eeq
The spin force is given by Eq. (\ref{papcoreqs1Iord}) with $U=U_{\rm (geo)}$ and should be nonzero to avoid trivial cases.
In the Schwarzschild spacetime the magnetic part of the Riemann tensor with respect to $U_{\rm (geo)}$ has the following nonvanishing frame components
\begin{eqnarray}
&&[{\mathcal H}(U_{\rm (geo)})^{\hat t\hat \theta},{\mathcal H}(U_{\rm (geo)})^{\hat t\hat \phi},{\mathcal H}(U_{\rm (geo)})^{\hat r\hat \theta},{\mathcal H}(U_{\rm (geo)})^{\hat r\hat \phi}]\nonumber\\
&&=-\frac{3M}{r^3}\gamma_{\rm (geo)}^2[\nu_{\rm (geo)}^{\hat r}\nu_{\rm (geo)}^{\hat \phi},
\nu_{\rm (geo)}^{\hat r}\nu_{\rm (geo)}^{\hat \theta},
-\nu_{\rm (geo)}^{\hat \phi},
\nu_{\rm (geo)}^{\hat \theta}]\,.
\end{eqnarray}
Clearly, one cannot take as the reference geodesic a purely radial one ($\nu_{\rm (geo)}^{\hat \theta}=0=\nu_{\rm (geo)}^{\hat \phi}$), since that leads to an identically vanishing ${\mathcal H}(U_{\rm (geo)})$.

Therefore, for the sake of an explicit example, let the reference world line be a circular geodesic in the equatorial plane at radius $r=r_0$. The associated  4-velocity is
\beq
\label{Ugeos}
U_{\rm (geo)}=\Gamma_K(\partial_t\pm\zeta_K\partial_\phi)=\gamma_K (e_{\hat t} \pm \nu_K e_{\hat \phi})\,,  
\eeq
where the simplified notation $U_{\rm (geo)}$ has been used in place of the more precise symbol $U_{\rm (geo,\pm)}$ for simplicity so one must keep in mind the sign correlation and
\beq\fl
\label{Ugeos2}
\Gamma_K=\frac{\gamma_K\nu_K}{r\zeta_K}\,, \
\zeta_K= \left( \frac{M}{r^3}\right)^{1/2} ,\
\nu_K=\left(\frac{M}{r-2M}\right)^{1/2}, \
\gamma_K=\left(\frac{r-2M}{r-3M}\right)^{1/2},
\eeq
and the timelike condition $\nu_K < 1$ is satisfied if $r>3M$.
The quantities $\gamma_K$, $\nu_K$, $\zeta_K$ are understood to be evaluated at $r=r_0$ and a positive (negative) sign corresponds to co-rotating (counter-rotating) orbits with respect to increasing values of the azimuthal coordinate $\phi$. 
The unit vector along the azimuthal direction in the local rest space of the circular geodesic which points in the direction of motion of the particle  (orthogonal to $U_{\rm (geo)}$ in the $t$-$\phi$ plane) is
\beq
\bar U_{\rm (geo)}=\gamma_K ( \pm e_{\hat \phi}+\nu_K e_{\hat t})\,,
\eeq
where the $\pm$ signs are correlated with those in $U_{\rm (geo)}$.

For a spinning particle, the  4-velocity $U$ 
\beq
\label{Uspin}
U=\gamma (e_{\hat t} +\vec \nu )
   \,, \quad 
\gamma=(1-\nu^2)^{-1/2}\,, \quad \nu=||\vec \nu||\,
\eeq
is parametrized by the spatial velocities $\nu^{\hat r}$, $\nu^{\hat \theta}$ and $\nu^{\hat \phi}$ 
\beq
\vec \nu =\nu^{\hat r}e_{\hat r}+\nu^{\hat \theta}e_{\hat \theta}+\nu^{\hat \phi}e_{\hat \phi}
  =  \tilde\nu^{\hat r}e_{\hat r}+ \tilde \nu^{\hat \theta}e_{\hat \theta}+(\pm\nu_K+\tilde\nu_{\hat \phi}) e_{\hat \phi}
\,.
\eeq
Here the spatial velocity deviations $\tilde\nu_{\hat r}=\nu_{\hat r}$, $\tilde\nu_{\hat \theta}=\nu_{\hat \theta}$ and $\tilde\nu_{\hat\phi}=\nu_{\hat\phi}-(\pm\nu_K)$ are understood to be small and all functions of them will be expanded to first order in these variables as well as in the spin variables $ S^\alpha$ below. For example, the gamma factor expanded in this way becomes
\beq
\gamma = \gamma_K (1 \pm \gamma_K^2 \nu_K \tilde\nu_{\hat\phi})
\,.
\eeq
The expansion of $U=U_{\rm (geo)}+({s}/{m})Y$ leads in turn to
\begin{eqnarray}
\label{Ydef}
U&=&U_{\rm (geo)}+\gamma_K\left[\tilde\nu_{\hat r}e_{\hat r}+\tilde\nu_{\hat \theta}e_{\hat \theta}+\left(\pm\gamma_K\tilde\nu_{\hat \phi}-\frac{\nu_K}{\gamma_K}\frac{\tilde r}{r_0}\right)\bar U_{\rm (geo)}\right]\,,
\end{eqnarray}
where the $\pm$ signs  are correlated with those of the geodesics, and the presence of the explicit geodesic gamma factors will simplify expressions below.
The second term on the right hand side of Eq. (\ref{Ydef}) is the velocity-deviation vector $Y$. 
Representing the path in terms of the background geodesic plus a deviation
$x^\alpha(\tau)=  x^\alpha_{\rm (geo)}(\tau) +\tilde x^\alpha(\tau)$
leads to the 4-velocity
\beq
  U^\alpha 
=  \frac{\rmd x^\alpha}{\rmd \tau} 
=  \frac{\rmd x^\alpha_{\rm (geo)}}{\rmd \tau} + \frac{\rmd \tilde x^\alpha}{\rmd \tau} \,,
\eeq
in terms of which one finds the linearized 4-velocity components to be
\begin{eqnarray} 
\label{motocoords}
 \frac{\rmd \tilde t}{\rmd \tau} &=& 
 \frac{\gamma_K \nu_K^2}{r_0 \zeta_K} \left(  -\frac{\nu_K}{r_0} \tilde r \pm \gamma_K^2 \tilde\nu_{\hat \phi} \right) \,,
\nonumber\\
 \frac{\rmd \tilde r}{\rmd \tau} &=&
  \frac{\gamma_K \zeta_K r_0}{ \nu_K} \tilde\nu_{\hat r} \,,
\nonumber\\
 \frac{\rmd \tilde \theta}{\rmd \tau} &=& 
  \frac{\gamma_K}{r_0} \tilde\nu_{\hat \theta} \,,
\nonumber\\
 \frac{\rmd \tilde \phi}{\rmd \tau} &=&
 \pm \frac{\gamma_K }{r_0} \left(  -\frac{\nu_K}{r_0} \tilde r \pm\gamma_K^2 \tilde\nu_{\hat \phi} \right)
=\pm\frac{\zeta_K}{\nu_K^2} \frac{\rmd \tilde t}{\rmd \tau}\,, 
\end{eqnarray}
and the circular geodesic is described by the equations
\beq\fl\qquad
t_{\rm (geo)} =  \Gamma_K \tau \,,\
r_{\rm (geo)} = r_0\,,\
\theta_{\rm (geo)} =\frac{\pi}{2}\,,\
\phi_{\rm (geo)} =  \pm\Omega_{\rm(orb)} \tau + \phi_0\,, 
\eeq
with
\beq
\Omega_{\rm(orb)}=\frac{\gamma_K \nu_K}{r_0}= \frac{1}{r_0}\sqrt{\frac{M}{r_0-3M}}
\,.
\eeq
The first order linear deviation velocities are thus given by
\begin{eqnarray} 
\label{nutildedef}
\tilde\nu_{\hat r}  =
  \frac{ \nu_K}{\gamma_K \zeta_K r_0}\frac{\rmd \tilde r}{\rmd \tau} \,,
\quad
  \tilde\nu_{\hat \theta} =
  \frac{r_0}{\gamma_K}\frac{\rmd \tilde \theta}{\rmd \tau} \,,
\quad
\tilde\nu_{\hat \phi} =\frac{1}{\gamma_K^2}\left(\frac{r_0}{\gamma_K}\frac{\rmd \tilde \phi}{\rmd \tau}
 \pm \frac{\nu_K }{r_0} \tilde r\right)\,, 
\end{eqnarray}
by inverting the relations (\ref{motocoords}).

The spin vector must be orthogonal to $U$ so to first order
\beq
\label{daTconds}
S^{\hat t}=S^{\hat r}\nu_{\hat r}+S^{\hat \theta}\nu_{\hat \theta}+S^{\hat \phi}\nu_{\hat \phi}=\pm\nu_KS^{\hat \phi}\,,
\eeq
which leads to the component expression
\beq
S = S^{\hat r}e_{\hat r} 
          +S^{\hat \theta}e_{\hat \theta}
          \pm \gamma_K^{-1} S^{\hat \phi}\bar U_{\rm(geo)}\,.
\eeq
This defines the spin components in the local rest frame of the circular geodesic.

Expanding the spin force defined in Eq.~(\ref{papcoreqs1}) to first order in the spatial velocity deviation vector leads to 
\beq
F^{\rm (spin)}=\mp3\gamma_K^2\nu_K\zeta_K^2 (S^{\hat \theta}e_{\hat r}+S^{\hat r}e_{\hat \theta})\,.
\eeq
It must equal $m$ times the spinning particle's acceleration whose components are
\begin{eqnarray}
\fl\quad
\label{aUcompts}
a(U)^t&=&\frac{\gamma_K^2\nu_K^2}{r_0}\left[\pm \frac{\gamma_K}{\zeta_K}\frac{\rmd\tilde\nu_{\hat \phi}}{\rmd \tau} +\tilde\nu_{\hat r} \right]
\equiv\frac{\nu_K}{r_0\zeta_K}a(U)^{\hat t}\,,\nonumber \\
\fl\quad
a(U)^r&=& \frac{r_0\zeta_K \gamma_K}{\nu_K}\frac{\rmd\tilde\nu_{\hat r}}{\rmd \tau}-\frac{\gamma_K^2\nu_K^2}{r_0^2}\tilde r \mp \frac{2 r_0\zeta_K^2 \gamma_K^2}{\nu_K}\tilde \nu_{\hat \phi}
\equiv\frac{r_0\zeta_K}{\nu_K}a(U)^{\hat r}\,,\nonumber \\
\fl\quad
a(U)^\theta&=&\frac{\gamma_K}{r_0}\frac{\rmd\tilde\nu_{\hat \theta}}{\rmd \tau}+\frac{\gamma_K^2\nu_K^2}{r_0^2}\tilde \theta
\equiv\frac{1}{r_0}a(U)^{\hat \theta}\,,\nonumber \\
\fl\quad
a(U)^\phi &=& \frac{\gamma_K^2}{r_0}\left[\gamma_K\frac{\rmd\tilde\nu_{\hat \phi}}{\rmd \tau}  \pm \zeta_K \tilde \nu_{\hat r}\right]
\equiv\frac{1}{r_0}a(U)^{\hat \phi}\,.
\end{eqnarray}
Solving for the first derivatives of the linear velocity corrections leads to the linearized equations 
\begin{eqnarray}
\label{motofin}
\frac{\rmd\tilde\nu_{\hat r}}{\rmd \tau}
&=&   \frac{\gamma_K\nu_K^3}{\zeta_K r_0^3} \tilde r
\pm2\gamma_K\zeta_K\left[\tilde\nu_{\hat \phi}
-\frac32\nu_K\zeta_K\frac{S^{\hat \theta}}{m}\right]
\,, \nonumber\\
\frac{\rmd\tilde\nu_{\hat \theta}}{\rmd \tau}
&=&  -\frac{\gamma_K \nu_K^2}{r_0} \tilde \theta
\mp3\zeta_K^2\gamma_K\nu_K\frac{S^{\hat r}}{m}
\,, \nonumber\\
\frac{\rmd\tilde\nu_{\hat \phi}}{\rmd \tau}
&=&
\mp\frac{\zeta_K}{\gamma_K}\tilde\nu_{\hat r}
\,.
\end{eqnarray}

These equations are a system of linear homogeneous second order ordinary differential equations for the coordinate deviations describing geodesic deviation to which the spin-curvature force contributes additional nonhomogeneous driving terms. By choosing initial conditions so that the 4-velocity is tangent to the circular geodesic
\beq
\tilde x^{\alpha}(0)=0=\frac{\rmd\tilde x^{\alpha}(0)}{\rmd\tau}  \,,
\eeq
one considers the solutions which describe deviations from that geodesic due to the spin-curvature force alone.

The spin evolution equations (\ref{papcoreqs2}) just represent parallel transport along the original circular geodesics, which leads to a simple rotation of the spin components $(S^{\hat \theta},\gamma_K^{-1} S^{\hat \phi})$ in the $r$-$\phi$ plane within the local rest space of the circular geodesics \cite{bcgj2008}. These equations are explicitly
\beq
\label{spineqs}
\frac{\rmd S^{\hat r}}{\rmd \tau}=\pm\frac{\zeta_K}{\gamma_K}S^{\hat \phi}\,, \qquad
\frac{\rmd S^{\hat \theta}}{\rmd \tau}=0\,, \qquad
\frac{\rmd S^{\hat \phi}}{\rmd \tau}=\mp\gamma_K\zeta_KS^{\hat r}\,.
\eeq
The complete set of equations of motion consists of the nine equations (\ref{nutildedef}), (\ref{motofin}), (\ref{spineqs}).

It is convenient to introduce a polar representation for the spin vector $S$
\beq\fl\quad
S^{\hat r}(\tau) =s_\Vert\cos\alpha(\tau)\,, \qquad
S^{\hat \phi}(\tau)=\gamma_Ks_\Vert\sin\alpha(\tau)\,, \qquad
S^{\hat \theta}=s_{\perp}\ ,
\eeq 
where $s_\perp$ is constant. 
We have then
\beq
\label{final_spin}
S=s_\Vert [\cos \alpha(\tau) e_{\hat r}\pm \sin \alpha(\tau) \bar U_{\rm (geo)}]+s_\perp e_{\hat \theta}
\equiv
 s_\Vert N_\Vert + s_\perp N_\perp\,,
\eeq
which defines two orthogonal unit vectors  orthogonal to $U_{\rm (geo)}$ which are parallely transported along the circular geodesic.
The properties of parallel transport along circular geodesics has been studied in detail in Ref. \cite{RJ}.
Then
Eqs.~(\ref{spineqs}) imply
\beq
\frac{\rmd s_{\Vert}}{\rmd \tau}=0\,, \qquad
\frac{\rmd \alpha}{\rmd \tau}=\mp\zeta_K\,,
\eeq
which leads to  $s_{\Vert}=const=[S^{\hat r}(0)^2+S^{\hat \phi}(0)^2/\gamma_K^2]^{1/2}$ and 
\beq
\label{sol_alpha}
\alpha(\tau)=\alpha_0\mp \zeta_K \tau \,.
\eeq
This represents a simple rotation of the components in the $r$-$\phi$ plane of the local rest space of the circular geodesic with the coordinate time geodesic frequency rather than the proper time frequency.

This result is a consequence of the properties of the Killing-Yano tensor which exists in the Schwarzschild spacetime
\beq
f=\frac12 f_{\mu\nu} \omega^\mu\wedge \omega^\nu=-r \omega^{\hat \theta}\wedge \omega^{\hat \phi}\,.
\eeq
Following a procedure outlined by Marck \cite{Marck1,Marck2} to complete a geodesic 4-velocity vector to a parallely propagated orthonormal frame along that geodesic, in our case one first obtains a vector $N_\perp=e_{\hat \theta}$ orthogonal to $U_{\rm (geo)}$ and parallely propagated along it in two steps:  contraction with the Killing tensor
\beq
f_{\hat \alpha \hat \beta}U_{\rm (geo)}^{\hat \beta}=\mp r_0\gamma_K \nu_K \delta^{\hat \theta}_{\hat \alpha}
\eeq
and then normalizing the result (modulo sign).
The remaining two orthonormal frame vectors orthogonal to both $U_{\rm (geo)}$ and $e_{\hat \theta}$ and parallely propagated along $U_{\rm (geo)}$ 
are determined by a single rotation angle $\alpha(\tau)$  in the $e_{\hat r}$-$\bar U_{\rm (geo)}$ plane, given their arbitrary initial values. The solution found above for $N_\Vert $ is such a vector.

\begin{figure}
\typeout{*** EPS figure 1}
\begin{center}
\includegraphics[scale=0.5]{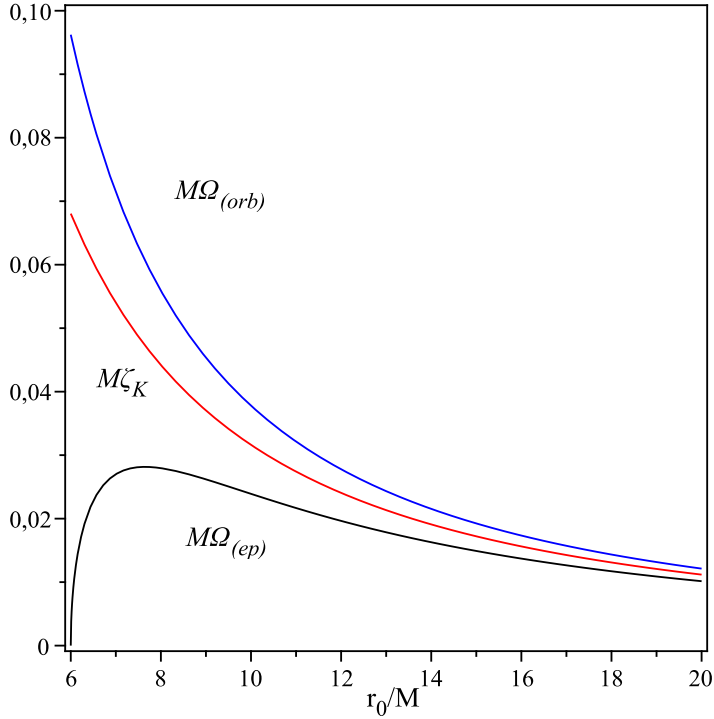}
\end{center}
\caption{The dimensionless angular velocities $M\Omega_{\rm (ep)}$, $M\Omega_{\rm (orb)}$ and $M\zeta_K$ are shown as functions of $r_0/M$.
The epicyclic frequency governing the radial perturbations of circular geodesics is real in the region $r_0>6M$, while the orbital frequency governing the geodesic oscillations out of the equatorial plane is real in the region $r_0>3M$ (see Eq.~(\ref{frequecies})).
}
\label{fig:1}
\end{figure}

Differentiating the first two velocity deviation equations (\ref{motofin}) with respect to $\tau$, and using the coordinate derivative relations (\ref{motocoords}), one obtains two decoupled second order differential equations for the velocity deviation variables $\tilde\nu^{\hat r},\tilde\nu^{\hat\theta}$ which define two frequencies for the solutions, and once $\tilde\nu^{\hat r}$ is found, the integration of the equation for  $\tilde\nu^{\hat\phi}$ immediately follows. These decoupled equations are
\beq
  \frac{\rmd^2 \tilde\nu^{\hat r}}{\rmd\tau^2} + \Omega_{\rm(ep)}^2 \tilde\nu^{\hat r} = 0\,,\quad
  \frac{\rmd^2 \tilde\nu^{\hat \theta}}{\rmd\tau^2} + \Omega_{\rm(orb)}^2 \tilde\nu^{\hat\theta} 
    = -3 \zeta_K^3 \nu_K \frac{S^{\hat \phi}}{m}
\,,
\eeq
where
\beq
\label{frequecies}
  \Omega_{\rm(ep)} \equiv \sqrt{\frac{M (r_0-6M)}{r_0^3 (r_0 -3M)}} \,,\qquad
  \Omega_{\rm(orb)} \equiv \frac{1}{r_0}\sqrt{\frac{M}{r_0-3M}}
\eeq
are respectively the well known epicyclic frequency governing the radial perturbations of circular geodesics and the orbital frequency governing the geodesic oscillations out of the equatorial plane (see Fig. \ref{fig:1}). The latter frequency together with the spin-precession frequency due to the spin oscillation driving term governs the polar angle oscillations about the equatorial plane.
Note that in order to have circular orbits at all, the condition $r_0>3M$ must be satisfied so $\Omega_{\rm(orb)}$ is always real and positive. However, in the interval $3M<r_0< 6M$ where the circular geodesic orbits are unstable, $\Omega_{\rm(ep)}^2$ is negative leading to hyperbolic functions for the homogeneous solutions of the second order differential equation for the radial velocity deviations even if the spin is set to zero, due to the geodesic instability alone. At $r_0= 6M$ where  $\Omega_{\rm(ep)}^2=0$, this leads to linear homogeneous solutions also reflecting the geodesic instability.
We therefore limit our attention to the zone $r_0>6M$. 


Instead of solving the decoupled second order equations for $\tilde\nu_{\hat r}$ and $\tilde\nu_{\hat\theta}$, one can integrate directly the second order equations for $\tilde t$ and $\tilde\phi$ once using the circular geodesic initial conditions, and retain the second order equations for $\tilde r$ and $\tilde\theta$, leading to the mixed system
\begin{eqnarray}\label{d2x}
\frac{\rmd \tilde t}{\rmd \tau}&=&-2\frac{\nu_K^2}{r_0\zeta_K}\Omega_{\rm(orb)}\tilde r\,,
\qquad
\frac{\rmd \tilde \phi}{\rmd \tau}=\mp 2 \Omega_{\rm(orb)}\frac{\tilde r}{r_0}\,,
\nonumber \\
\frac{\rmd^2 \tilde r}{\rmd \tau^2}&=& -\Omega_{\rm(ep)}^2 \tilde r\mp A_\perp \sigma_\perp\,,\qquad
\frac{\rmd^2 \tilde \theta}{\rmd \tau^2}= -\Omega_{\rm(orb)}^2 \tilde \theta \mp A_\Vert \sigma_\Vert \cos \alpha(\tau)\,,
\end{eqnarray}
where $A_\perp$ and $A_\Vert$ are constant
\beq
A_\perp=3\gamma_K^2 \left(\frac{M}{r_0}\right)^2\zeta_K\,,\qquad 
A_\Vert=3\gamma_K^2 \left(\frac{M}{r_0}\right)^2\frac{\nu_K}{r_0^2}\,,
\eeq
and  the dimensionless spin quantities
\beq
\sigma_\Vert=\frac{s_\Vert}{mM}  \,,\qquad 
\sigma_\perp=\frac{s_\perp}{mM}\,
\eeq
have been introduced.
Note that the first pair of equations imply
\beq
\frac{\rmd \tilde \phi}{\rmd \tau}=\pm \frac{\zeta_K}{\nu_K^2}\frac{\rmd \tilde t}{\rmd \tau}\,,
\eeq
which with the above initial conditions gives
\beq
\tilde t=\pm (\nu_K^2/\zeta_K)\tilde\phi\,.
\eeq
Thus one finds the solution
\begin{eqnarray}
\fl\qquad
\tilde r
&=& \pm r_0\Sigma_\perp[1-\cos(\Omega_{\rm(ep)}\tau)]
\,, \nonumber\\ \fl\qquad
\tilde\theta 
&=&  
\Sigma_\Vert\left[\frac{\zeta_K}{\Omega_{\rm(orb)}} \sin \alpha_0 \sin(\Omega_{\rm(orb)}\tau) \pm  \cos \alpha_0  \cos(\Omega_{\rm(orb)}\tau)\mp \cos(\alpha(\tau))\right]\,,
\nonumber \\
\fl\qquad
\tilde\phi
&=&
2\frac{\Omega_{\rm(orb)}}{\Omega_{\rm(ep)}}
\Sigma_\perp[\sin(\Omega_{\rm(ep)}\tau) - \Omega_{\rm(ep)}\tau]
\,,
\end{eqnarray}
with the following convenient amplitude factors
\beq
\Sigma_\perp = -\frac{A_\perp}{r_0\Omega_{\rm(ep)}^2} \sigma_\perp\,, \qquad
\Sigma_\Vert = \frac{M}{r_0}\frac1{\nu_K} \sigma_{\Vert}\,.
\eeq
The coefficients in the amplitudes of the oscillations contain the factors
\beq
     \frac{\zeta_K}{\Omega_{\rm(orb)}} = \left(1-\frac{3M}{r_0}\right)^{1/2}\,,\quad
      \frac{\Omega_{\rm(orb)}}{\Omega_{\rm(ep)}} = \left(1-\frac{6M}{r_0}\right)^{-1/2}\,.
\eeq
Note that both $\Sigma_\perp$ and this second factor become infinite as one approaches $r_0=6M$, which increases the size of the perturbation beyond its validity, so one must restrict $r_0$ to be a bit bigger than this value to avoid consistency problems. This effect stretches out the amplitudes of the horizontal oscillations.

Finally observe that the radial oscillations are centered about the new equilibrium radius $r=r_0(1\pm \Sigma_\perp)$, where the radial gravitational and spin-curvature forces balance, inducing a slightly different azimuthal velocity to ensure this balance. The first of Eqs. (\ref{motofin}) shows how for $\tilde\nu_{\hat r}=0$ at this radial equilibrium, the constant value of the azimuthal velocity must adjust to guarantee this equilibrium. However, since we choose to examine the solutions with the initial value $\tilde r=0$, this equilibrium itself is not part of our family of solutions.

The first order velocities  $\tilde\nu_{\hat r}$, $\tilde\nu_{\hat \phi}$ and $\tilde\nu_{\hat \theta}$ follow from Eq. (\ref{nutildedef}). 
The solution for $U$ (which at $\tau=0$ is aligned with the circular geodesic at $r_0$) is then given by
\begin{eqnarray}\fl\label{Usol}
U&=&U_{\rm (geo)}\pm  \nu_K \Sigma_\perp \left[\frac{\Omega_{\rm(ep)}}{\zeta_K}\sin (\Omega_{\rm(ep)}\tau) e_{\hat r}+2[\cos(\Omega_{\rm(ep)}\tau)-1]\bar U_{\rm (geo)}\right]\nonumber\\
\fl
&&+ r_0\zeta_K \Sigma_\Vert \left[\sin \alpha_0 \cos(\Omega_{\rm(orb)}\tau) \mp  \frac{\Omega_{\rm(orb)}}{\zeta_K}\cos \alpha_0  \sin(\Omega_{\rm(orb)}\tau)\right.\nonumber \\
\fl
&&
\left. -\sin(\alpha(\tau))\right]e_{\hat \theta}\,. 
\end{eqnarray} 

These solutions have a small oscillating radial motion compared to the circular geodesics at $r=r_0$
\beq
r(\tau)=r_0 \left[1 \pm \Sigma_\perp (1-\cos\Omega_{\rm(ep)}\tau)\right]\,,
\eeq
representing an oscillation about the slightly adjusted equilibrium radius $r_0(1\pm \Sigma_\perp)$ with an amplitude $r_0| \Sigma_\perp|$, i.e., in a strip of width $2r_0 |\Sigma_\perp|$ outside or inside the geodesic radius.
If $\pm\Sigma_\perp >0$ so the orbital angular velocity and $\Sigma_\perp$ have the same sign, this oscillation is outside $r_0$ at a slightly smaller azimuthal speed (see Eq. (\ref{Usol})), but if  $\pm\Sigma_\perp <0$ so that they have the opposite sign, it is inside $r_0$ at a slightly larger azimuthal speed (spin-orbit coupling).

\begin{figure}
\typeout{*** EPS figure 2}
\begin{center}
\includegraphics[scale=0.4]{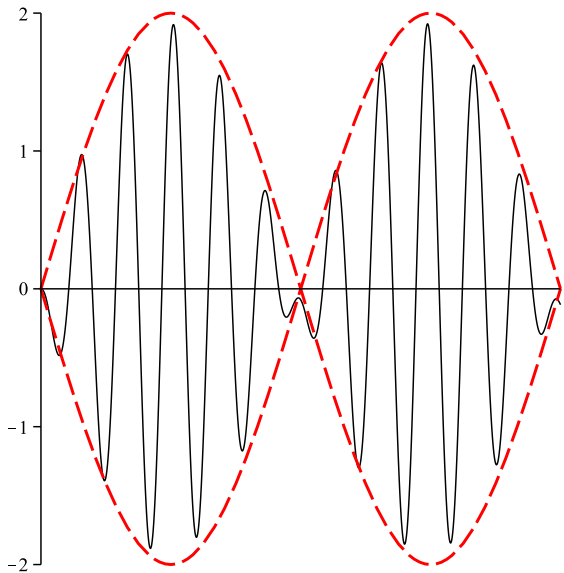}
\end{center}
\caption{The behavior of the first order correction $\tilde\theta$ to the polar motion in units of $\Sigma_\Vert$ is shown as a function of the proper time for the choice of parameters $r_0/M=10$ and $\alpha_0=\pi/4$ in the case of a co-rotating orbit, together with the envelope of the beating function (the second term inside the square brackets of Eq.~(\ref{thfin2})), also showing the small contribution of the nonbeating term in comparison.
}
\label{fig:2}
\end{figure}

\begin{figure}
\typeout{*** EPS figure 3}
\begin{center}
\includegraphics[scale=0.4]{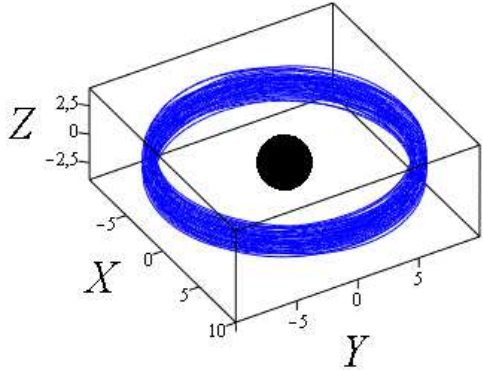}
\end{center}
\caption{
The corotating orbit inside the circular geodesic at $r_0=10M$, for exaggerated values $\Sigma_\Vert=0.25=\Sigma_\perp$ of the two spin components, showing the band into which the original orbit spreads due to the spin.
}
\label{fig:3}
\end{figure}

The polar angle motion about the equatorial plane 
\beq\fl
\label{thfin}
\theta(\tau)=\frac{\pi}{2}+\Sigma_\Vert \left[\frac{\zeta_K}{\Omega_{\rm(orb)}}\sin \alpha_0 \sin(\Omega_{\rm(orb)}\tau) \pm  \cos \alpha_0  \cos(\Omega_{\rm(orb)}\tau)\mp\cos(\alpha(\tau))\right]
\eeq
consists of a superposition of two frequencies: $\Omega_{\rm(orb)}$ which appears explicitly and $\zeta_K$ which appears in $\alpha(\tau)$, two frequencies associated with the orbital motion which are approximately equal even for moderate values of $r_0/M>6$. This leads to a beat phenomenon as already pointed out in \cite{mashsingh} for the more complicated Kerr case: the
larger (average) frequency harmonic oscillation in the polar angle deviation approximately at the Keplerian frequency is modulated by the much smaller (half-difference) beat frequency, in terms of which Eq. (\ref{thfin}) can be rewritten as 
\begin{eqnarray}
\label{thfin2}
\theta(\tau)&=&\frac{\pi}{2}+\Sigma_\Vert \left[\left(\frac{\zeta_K}{\Omega_{\rm(orb)}}-1\right)\sin \alpha_0 \sin(\Omega_{\rm(orb)}\tau)\right.
\nonumber\\
&&\left.+2\sin\left(\alpha_0 \mp\frac{\Omega_{\rm(orb)}+\zeta_K}{2}\tau\right)\sin\left(\frac{\Omega_{\rm(orb)}-\zeta_K}{2}\tau\right)\right]\,.
\end{eqnarray}
The last term is responsible for the beat (see Fig. \ref{fig:2}), with beat frequency $(\Omega_{\rm(orb)}-\zeta_K)/2$, which for $r_0\gg 2M$ has the approximate value $3M\zeta_K/4 $ as in the Kerr case \cite{mashsingh}.
The first term is typically much smaller than the beating term even for $r_0\approx7M$.
However, from Eq. (\ref{thfin2}) using the triangle inequality one finds
\beq
\left|\theta(\tau)-\frac{\pi}{2}\right|\le\Sigma_\Vert\left(3-\frac{\zeta_K}{\Omega_{\rm(orb)}}\right)\,,
\eeq
which is characterized roughly by the interval width $2\Sigma_\Vert$.
For $r_0\gg 2M$ where $|\zeta_K/\Omega_{\rm(orb)}|\approx1$ then the arclength amplitude of the oscillation is roughly $2r_0\Sigma_\Vert$.
For example, at $r_0=10M$, the interval of arclength is a band characterized by the approximate amplitude $2r_0\Sigma_\Vert $, while the radial band has roughly a thickness of $2r_0|\Sigma_\perp| $. This leads to the orbit filling a rectangular cross-section tube in Fig. \ref{fig:3} elongated by a factor of 2 in the vertical direction for the equal values chosen for the two independent rescaled spin components $\Sigma_\Vert $ and $|\Sigma_\perp| $, with values exaggerated to show the tube.


Finally, the azimuthal motion 
\beq
\phi(\tau)=\phi_0\pm\Omega_{\rm(orb)}\tau-2\frac{\Omega_{\rm(orb)}}{\Omega_{\rm(ep)}}
\Sigma_\perp[\sin(\Omega_{\rm(ep)}\tau) - \Omega_{\rm(ep)}\tau]
\eeq
oscillates around the geodesic value with the same frequency characterizing the radial motion,  apart from a secular increase in that angle.
The corrected Keplerian angular velocity due to the spin is
\beq
\zeta\equiv
\frac{\rmd \phi}{\rmd t}=\pm \zeta_K +2\Sigma_\perp\frac{\zeta_K}{\gamma_K^2} \left[\cos \left(\frac{\Omega_{\rm(ep)} t}{\Gamma_K}\right)-1\right]\,,
\eeq
whose average value is
\beq
\zeta_{\rm(avg)}=\pm \zeta_K -2\Sigma_\perp\frac{\zeta_K}{\gamma_K^2}\,.
\eeq
For $r_0\gg 2M$ where $|\Omega_{\rm(orb)}/\Omega_{\rm(ep)}|\approx1$, then the amplitude of the azimuthal oscillations is roughly $2|\Sigma_\perp|$ corresponding to an arclength $2r_0|\Sigma_\perp| $ agreeing with the radial oscillation amplitude.

Let us examine the following special cases.

\begin{enumerate}

\item{$s_\Vert=0$ ($\sigma_\Vert=0=\Sigma_\Vert$)}.

In this case $S^{\hat r}=0=S^{\hat \phi}$ and $\tilde\theta=0$.
Therefore, the motion is confined to the equatorial plane at the epicyclic frequency and the spin vector has only the constant $\theta$ component, i.e., the particle can only have spin up or down. The radial oscillations are either inside or outside the circular orbit radius in a strip each roughly of width $2r_0|\Sigma_\perp|$ on each side, with a similar arclength amplitude in the azimuthal direction, apart from a secular increase in that angle which occurs at slightly different azimuthal speed inside and outside the radius of the circular geodesic.

\item{$s_\perp=0$ ($\sigma_\perp=0=\Sigma_\perp$)}.

In this case $S^{\hat \theta}=0$ and $\tilde r=0=\tilde\phi$ and 
the motion is confined to the same radius and azimuthal angle as the corresponding circular geodesic, oscillating vertically above and below it with an arclength amplitude of roughly $2r_0\Sigma_\Vert$. The spin vector rotates within the equatorial plane.

\end{enumerate}

In the case of a general orientation of the spin vector, these two approximate amplitudes $2r_0|\Sigma_\perp|$  and $2r_0\Sigma_\Vert$ characterize the independent motion in the horizontal and vertical directions respectively.
For a fixed total spin $s$ but different orientations of the horizontal  and vertical spin vector components, the deviation from the circular geodesic would be roughly confined to an ellipsoid of revolution with an elliptical vertical cross-section, if it were not for the varying azimuthal velocities which depend on the spin orientation. This latter effect spreads out the trajectories azimuthally for different spin orientations.
 If we introduce the constant spin polar angle by
$ s_\perp=s\cos\beta, s_\Vert=s\sin\beta$, then different orbits for the same $s$ but different $\beta$ and initial value $\alpha(0)$ of the azimuthal spin angle 
will lie roughly within an ellipse in the plane perpendicular to the motion whose equation follows from
the total spin relation
\beq
1= \frac{\Sigma_\perp^2}{\mathcal{A}^2} + \frac{\Sigma_\Vert^2}{\mathcal{B}^2}\,,
\eeq
which is an ellipse with semi-axes satisfying
\begin{eqnarray}
  \mathcal{B} &=& \frac{s}{mr_0\nu_K}\,,\nonumber\\
  \mathcal{A} &=& 3 \frac{r_0-2M}{r_0-6M} \left(\frac{r_0}{r_0-2M} \right)^{1/2} \left(\frac{M}{r_0} \right)
 \mathcal{B}\,,\nonumber\\
&&\lim_{r_0\gg2M} \frac{\mathcal{A} }{\mathcal{B} } =\frac{3M}{r_0} \to0\,,
\lim_{r_0\to6M} \frac{\mathcal{A} }{\mathcal{B} } =\infty
\,.
\end{eqnarray}

In the general case,
the three spatial deviations are characterized by motion within intervals of arclength $2r_0|\Sigma_\perp|$ (radial) and $2r_0\Sigma_\Vert$ (angular) away from the circular geodesic but at different phases, so the solution curves of fixed total spin wander roughly within an ellipsoid about the corresponding geodesic position with these semiaxes, with an additional spin-orientation dependent secular drift in the azimuthal angle which spreads out the ellipse azimuthally in a shearing motion.
At large radii the semi-axis ratio $\mathcal{A} /\mathcal{B} $ gets very small, squashing the ellipse flat in the horizontal direction, with the corresponding tube of orbits degenerating into a vertical strip. On the other hand approaching $r_0=6M$, the ellipsoid is stretched horizontally since $\mathcal{A}\to0$, leading to large horizontal oscillations.

\section{Conclusions}

We have studied the deviation of a spinning particle moving in the Schwarz\-schild spacetime compared to a circular geodesic in the equatorial plane, assuming that both orbits have the same initial 4-velocity,
 in order to focus on how the presence of the spin changes that geodesic motion.
 Within the Mathisson-Papapetrou-Dixon model and linearizing the equations in the spin parameter, assumed to be very small as required in order to ignore the back reaction on the spacetime geometry, we have explicitly evaluated the deviation vector.
We have also shown that during its motion the orbit of the spinning particle has a radial component which oscillates in a circular ring either inside or outside the geodesic radius depending on the relative sign of the vertical component of the spin and the orbital velocity.
The azimuthal motion also oscillates around the geodesic value with the same frequency characterizing the radial motion, apart from a secular drift which occurs at slightly different speeds for the inner and outer radial oscillations.
The $\theta$ motion is instead characterized by the superposition of two frequencies, one the same as in both the horizontal motion, and the other governing the oscillations of the spin vector;
the result is a harmonic oscillation at  the orbital frequency, modulated by a beat frequency.
Thus the orbit is confined to a tube inside or outside the initial circular orbit at the radius $r_0$.

If the spin of the particle is initially aligned along the direction perpendicular to the equatorial plane ($s_\Vert=0$) the tube collapses to a flat (horizontal) circular corona.
If the spin is initially horizontal ($s_\perp=0$), then the tube collapses to a flat vertical band. 
 
Finally, an interesting consequence of this analysis is that the spin of the test particle leads to a small spread in its position compared to a spinless particle following the same circular geodesic. In other words without precise knowledge of the initial orientation of the spin when the test particle has the initial data of the corresponding circular geodesic, it can be found somewhere within a small neighborhood of that geodesic within the limits to the oscillating deviations in the three independent directions, modulo a small secular drift in the azimuthal direction which stretches out the two halves of this ellipsoid in opposite azimuthal directions.

\end{document}